\documentclass[11pt,a4paper]{article}
\begin{document}
\thispagestyle{empty}
\noindent
{\Large \textbf{A realist view on the treatment of identical particles}}

\bigskip
\bigskip
\noindent
{\textbf{Arthur Jabs}}

\bigskip
\noindent
Alumnus, Technische Universit\"at Berlin

\noindent
Vo\ss str.~9, 10117 Berlin, Germany
\\e-mail: arthur.jabs@alumni.tu-berlin.de
\bigskip
\\(17 January 2015)
\bigskip
\smallskip

\newcommand{\rmi}{\mathrm{i}}
\newcommand{\bfitr}{\emph{\boldmath $r$}}
\newcommand{\Pal}{P_{\alpha}}
\newcommand{\ud}{\mathrm{d}}
\newcommand{\page}[1]{\hfill \mbox{{#1}\hspace{13 em}}\\} 

\noindent
\textbf{Abstract.}
Some basic concepts concerning systems of identical particles are discussed in the framework of a realist interpretation, where the wave function \emph{is} the quantum object and $|\psi(\bfitr)|^2\ud^3 r$ is the probability that the wave function causes an effect about the point $\bfitr$. The topics discussed include the role of Hilbert-space labels, wave-function variables and wave-function parameters, the distinction between permutation and renaming, the reason for symmetrizing the wave function, the reason for antisymmetric wave functions, the spin-statistics theorem, the correction term $-k\ln N!$ in the entropy, the Boltzmann limit of the Fermi and Bose cases, and the derivation of the Fermi and Bose distributions.
\\
\\
\rule{\textwidth}{.3 pt}
\newcommand{\leer}{\hspace*{\fill}}
\newcommand{\lll}{\hspace*{90pt}}
\newcommand{\hsp}{\hspace*{60pt}}
\begin{center}
\vspace{-20pt}
\noindent
\hsp 1~~Introduction \leer 1\hsp \lll \\
\hsp2~~The realist interpretation  \leer 2\hsp \lll \\ 
\hsp3~~Wavepackets   \leer 3\hsp \lll \\
\hsp4~~Hilbert-space labels for non-identical particles  \leer 4\hsp \lll \\ 
\hsp5~~Self interactions   \leer 7\hsp \lll \\ 
\hsp6~~Renaming and permutation symmetry  \leer 7\hsp \lll  \\
\hsp7~~Identical particles. Basic features  \leer 9\hsp \lll \\
\hsp8~~What is permuted?   \leer 11\hsp \lll \\
\hsp9~~Symmetric operators  \leer 13\hsp \lll \\ 
\hsp10 Symmetric wave functions  \leer 14\hsp \lll \\
\hsp11 Antisymmetric wave functions   \leer 16\hsp \lll \\
\hsp12 The term $-k\ln N!$ in the entropy   \leer 18\hsp \lll \\
\hsp13 Bose, Fermi and Boltzmann probabilities  \leer 18\hsp \lll \\ 
\hsp14 The Bose and Fermi distributions   \leer 21\hsp \lll \\
\hsp References   \leer 22-24\hsp \lll \\
\end{center}
\vspace{-10pt}
\rule{\textwidth}{.3 pt}
\\
\\
\noindent 
{\textbf{1~~Introduction}}
\smallskip

\noindent
In this note I want to show that quantum mechanics, and in particular quantum mechanics of identical particles, becomes more appealing when formulated in the language of epistemological realism [1], [2]. The discussed topics can be seen from the above-listed section headings. The mathematical level is elementary. All formulas, except one, are taken from standard textbooks. In fact, there is hardly anything new (not yet published), but things are composed in a different way and seen from a different point of view.

\bigskip
\noindent
{\textbf{2~~The realist interpretation}}
\smallskip

\noindent
Realist interpretation in the present note always means the interpretation expounded in [1]. This interpretation is opposed to the Copenhagen interpretation. Only the basic features will be outlined here as they are needed for the topics discussed in the present paper. [1] gives the full treatment.

The realist interpretation derives its name from the feature that it considers the quantum objects to have their properties, whichever these are, independent of whether we observe them or not. This is of course not a matter to be proved or disproved. It is a manner of speaking, a way we express ourselves in our everyday language, and it is in fact a point of view developed by children at the age of about one year and a half [3]. This type of realism is not in conflict with the philosophical constructivism of von Glasersfeld and von Foerster [4].	

A basic feature of the considered realist interpretation is that a pointlike position is never among the properties of a quantum object. Rather these objects are extended. They are represented in the mathematical formalism by wavepackets of (Schr\"odinger) wave functions. The wavepacket
 $\psi(\bfitr)$ 
is not the probability amplitude of the behaviour of an ensemble of particles but is a real objective physical field, like a pulse of an electromagnetic wave. There is no wave-particle duality. The concept of the particle as the carrier of the pointlike effects is as dispensible as the ether was as a carrier of the electromagnetic waves. The size of the ``particle'' is just the extension of the wavepacket. The point particles of the Copenhagen interpretation are like the emperor's new clothes in Andersen's fairy tale, where it is the child who reveals: ``But he hasn't got anything on!'' [5].

The expression $|\psi(\bfitr)|^2\ud^3 r$ is, in appropriate physical situations, numerically equal to the probability that the wavepacket causes a physical effect in $\ud^3 r$ about the point $\bfitr$, whether detected or not. The quantum object is like a cloud triggering thunder and lightning here and there, and $|\psi|^2\ud^3 r$ may be called an action probability.
In this action the wavepacket acts as a whole. This is an additional property the wavepacket is endowed with, in order to take the quantum effects proper into account. It means that the wavepacket is an elementary region of space, representing one field quantum = one particle (or an integral number of quanta, see below). This also means that there are no dynamical self interactions.

Identical as well as non-identical wavepackets are not always independent of each other but may become entangled when they overlap at some time in space. This is a manifestation of the fact that the wave function basically is a function in multiparticle configuration space, which entails nonlocal effects.

Wavepackets representing Bose particles may \emph{condense} and form one single wavepacket, comparable to Einstein condensation. This wavepacket then represents an integral number of quanta. The condensation is conceived as a real physical process, but in no case does it mean that there are point particles that come to lie in one and the same wave function, like balls in a bag.

Actually, it would be preferable to use the word ``similar'' in place of ``identical'' (like in Dirac's book [6]) because the wavepackets representing ``identical particles'' may still have different shapes. It would also be preferable to speak of ``wavepacket'' or ``quantum wavepacket'' instead of ``particle'' with its rather misleading connotations. I nevertheless continue to use the word ``identical particles'', following entrenched use, but the reader should notice that I always mean similar wavepackets.

\bigskip
\noindent
{\textbf{3~~Wavepackets}}
\smallskip

\noindent
In order to have a concrete example we consider a one-dimensional wavepacket of Gaussian form, meant to represent a free particle (cf. [7, p.~64] with the replacements $x \rightarrow x-x_0,\; t \rightarrow t-t_0,\; m \rightarrow m_0,\; a \rightarrow \sigma)$
\begin{displaymath}
\psi(m_0,\sigma,x_0,t_0,k_0,x,t) =  \left(\frac{2}{\pi \sigma^2(1+A^2(t))}\right)^{1/4}\times \exp\!\left[-\frac{\left(x-x_0-\frac{\hbar k_0}{m_0}(t-t_0)\,\right)^2}{\sigma^2(1+A^2(t))}\right]
\end{displaymath}
\begin{equation}
\times
\exp\;\rmi\!\left\{ 
\frac{A(t)\left(x-x_0-\frac{\hbar k_0}{m_0}(t-t_0)\right)^2}{\sigma^2(1+A^2(t))} 
-\mbox{$\frac{1}{2}$}\arctan A(t) +k_0(x-x_0)- \mbox{$\frac{\hbar k_0^2}{2m_0}$}(t-t_0) 
\right\}
\end{equation}
with
\begin{equation}
A(t)=\frac{2\hbar}{m_0\sigma^2}(t-t_0),
\end{equation}
where $m_0$ = mass, $\sigma$ = full width of the wavepacket ($|\psi|^2$) at $t=t_0$, $x_0$ = average position (=position of the centre) at $t=t_0$, $t_0$ = time at which the packet has its minimum width, $k_0$= average (=centre) momentum of the wavepacket ($x_0$ and $k_0$ being like initial conditions), $x$, $t$ = space and time variables of the wavepacket. The maximum (=centre) of this packet is at $x=x_0+\frac{\hbar k_0}{m_0}(t-t_0)$. It moves with velocity $\frac{\hbar k_0}{m_0}$ in positive $x$-direction. $A(t)$ describes the spreading.

The arguments of $\psi$ on the left-hand side of Eq.~(1) will be called state \emph{quantities}. These are of two types: \emph{function parameters} ($m_0,\sigma,x_0,t_0,k_0$) and \emph{function variables} ($x,t$). The wave function $\psi$ is a function of the variables $x$ and $t$, and the shape and position of this function is determined by the parameters $m_0, \sigma, x_0, t_0$, and $k_0$. The parameters in turn are of two types: \emph{intrinsic} and \emph{external}.

The intrinsic parameters denote those permanent quantities that define the particular kind of particles. In (1) it is the mass $m_0$, and in more general cases also the charge, total spin, parity etc. Particles of the same kind are considered identical.

The external parameters (here $\sigma, x_0, t_0, k_0$) denote those quantities whose values may differ among the particles of the same kind. That is, even if the external parameters are different in two particles, the particles may be considered identical (realist: the wavepackets may be considered similar). In our case the external parameters are $\sigma, x_0, t_0$, and $k_0$, and in Sections 8 and  11 below they will be complemented by the spin component $m$ and the azimuthal spinor angle $\chi$. Also, the considered single-particle wave function need not be an eigenfunction of a complete set of commuting observables. In the general case it is an expansion in terms of such basic functions. The expansion coefficients are then also counted among the external parameters of the wave function.

In the Copenhagen interpretation the intrinsic parameters are ascribed to the (point) particles (cf. [8]) and the external parameters to the wave function. In our realist interpretation there is only a wavepacket, and this carries all parameters, intrinsic and external.

For later reference we note some alternative notations for the wave function
\linebreak
$\psi(m_0,\sigma,x_0,t_0,k_0,x,t)$. We may write
\begin{equation}
\psi(m_0,u,x,t),
\end{equation}
where we have put all external function (state) parameters into the symbol $u$. Or we may write
\begin{equation}
\psi_{m_0,u}(x,t),
\end{equation}
where we have put the parameters as labels at the function symbol $\psi$. Or even
\begin{equation}
\varphi(x,t),
\end{equation}
where we have combined the function symbol $\psi$ with its defining parameters $m_0, u$ into the letter $\varphi$.

\bigskip
\noindent
{\textbf{4~~Hilbert-space labels for non-identical particles}}
\smallskip

\noindent
In order to understand the treatment of identical particles in the current formalism it is necessary first to understand the treatment of non-identical particles. In this treatment Hilbert-space labels are introduced to distinguish the particles. A priori one would expect that here Hilbert-space labels are superfluous because non-identical particles can already be distinguished by their different intrinsic parameters like mass, charge etc. Nevertheless, in the current formalism of quantum mechanics Hilbert-space labels are employed even when non-identical particles are treated. Why?

Consider a system of an electron and a proton interacting via the Coulomb force. The Schr\"odinger equation for this system reads [7, p. 788], [9, p. 89, 90], [10, p. 128, 129, 411], 
\begin{displaymath}
\rmi \hbar \frac{\partial}{\partial t}\Psi(x_1,y_1,z_1,x_2,y_2,z_2,t)=
\end{displaymath}
\begin{displaymath}
=-\left[\frac{\hbar^2}{2m_{01}}\left(\frac{\partial^2}{\partial x_1^2}+\frac{\partial^2}{\partial y_1^2}+\frac{\partial^2}{\partial z_1^2}\right) 
+\frac{\hbar^2}{2m_{02}}\left(\frac{\partial^2}{\partial x_2^2}+\frac{\partial^2}{\partial y_2^2}+\frac{\partial^2}{\partial z_2^2}\right)\right.
\end{displaymath}
\begin{equation}
+\left. \frac{e^2}{\sqrt{(x_1-x_2)^2+(y_1-y_2)^2+(z_1-z_2)^2}}\right]
\Psi(x_1,y_1,z_1,x_2,y_2,z_2,t).
\end{equation}
The indices at the variables $x,y,z$ distinguish the proton variables from the electron variables. These indices are usually called particle labels, but as in our realist interpretation the ``particles'' \emph{are} the wavepackets we shall call these labels just \emph{Hilbert-space labels}, following a suggestion by Dieks [11]:

\smallskip
\centerline{``particle labels'' $\rightarrow$ ``Hilbert-space labels''.}

\smallskip
\noindent
These Hilbert-space labels will always be put in parentheses. They have nothing to do with the shape and position of the wavepackets.

\newcommand{\qf}{\frac{\hbar^2}{2m_{01}}}
\newcommand{\qs}{\frac{\hbar^2}{2m_{02}}}

For our purposes it suffices to consider the $x$-dependence only and to write Eq.~(6) in the form
\begin{displaymath}
\rmi \hbar \frac{\partial}{\partial t}\Psi(x^{(1)},x^{(2)},t)
\end{displaymath}
\begin{equation}
=-\left[\frac{\hbar^2}{2m_{01}}\frac{\partial^2}{\partial(x^{(1)})^2}+\frac{\hbar^2}{2m_{02}}\frac{\partial^2}{\partial(x^{(2)})^2}+\frac{e^2}{|x^{(1)}-x^{(2)}|} \right] \Psi(x^{(1)},x^{(2)},t).
\end{equation}
Let us try to do without the Hilbert-space labels. Consider first the still simpler case with no interaction and with a wave function that is a product of two functions like those of Eq.~(3)
\begin{equation}
\Psi(x^{(1)},x^{(2)},t)=\psi(m_{01},u_1,x^{(1)},t)\;\psi(m_{02},u_2,x^{(2)},t).
\end{equation}
In contrast to the Hilbert-space labels the function parameters $m_{01},u_1$ and $m_{02},u_2$ do determine the mathematical form of the wave functions: $\psi(m_{01},u_1,x,t)$ has a different shape and position than $\psi(m_{02},u_2,x,t)$, independent of whether it refers to the proton or to the electron, that is, independent of whether $x=x^{(1)}$ or $x=x^{(2)}$.

Insert function~(8) into the Schr\"odinger equation (7) without the Coulomb interaction term:
\begin{displaymath}
-\rmi\hbar\frac{\partial}{\partial t}\psi(m_{01},u_1,x^{(1)},t)\;\psi(m_{02},u_2,x^{(2)},t)
\end{displaymath}
\begin{equation}
=\left[\frac{\hbar^2}{2m_{01}}\frac{\partial^2}{\partial(x^{(1)})^2}+\frac{\hbar^2}{2m_{02}}\frac{\partial^2}{\partial(x^{(2)})^2}\right]\psi(m_{01},u_1,x^{(1)},t)\;\psi(m_{02},u_2,x^{(2)},t).
\end{equation}
The right-hand side can be written as
\begin{displaymath}
\psi(m_{02},u_2,x^{(2)},t)\frac{\hbar^2}{2m_{01}}\frac{\partial^2}{\partial(x^{(1)})^2}\psi(m_{01},u_1,x^{(1)},t)
\end{displaymath}
\begin{equation}
+\psi(m_{01},u_1,x^{(1)},t)\frac{\hbar^2}{2m_{02}}\frac{\partial^2}{\partial(x^{(2)})^2}\psi(m_{02},u_2,x^{(2)},t).
\end{equation}
Now compare this with the case where all Hilbert-space labels are removed and the usual rules of calculus are applied. The left-hand side remains the same, except that the Hilbert-space labels are missing, but the right-hand side becomes
\begin{displaymath}
\left[\frac{\hbar^2}{2m_{01}}+\frac{\hbar^2}{2m_{02}}\right]\frac{\partial^2}{\partial x^2}\psi(m_{01},u_1,x,t)\psi(m_{02},u_2,x,t)
\end{displaymath}
\begin{displaymath}
=\psi(m_{02},u_2,x,t)\left[\qf+\qs\right]\frac{\partial^2}{\partial x^2}\psi(m_{01},u_1,x,t)
\end{displaymath}
\begin{displaymath}
+\psi(m_{01},u_1,x,t)\left[\qf+\qs\right]\frac{\partial^2}{\partial x^2}\psi(m_{02},u_2,x,t)
\end{displaymath}
\begin{equation}
+2\left[\qf+\qs\right]\left(\frac{\partial}{\partial x}\psi(m_{01},u_1,x,t)\right)\left(\frac{\partial}{\partial x}\psi(m_{02},u_2,x,t)\right).
\end{equation}
This is different from (10) because each operator $\frac{\hbar^2}{2m_{0i}}\frac{\partial^2}{\partial(x^{(i)})^2}$ ($i$=1, 2) now operates on both $·\psi(m_{01},u_1,x,t)$ and $\psi(m_{02},u_2,x,t)$, that is, the operators operate no longer only in their own associated Hilbert space but encroach the other Hilbert space too. So, simply removing the Hilbert-space labels does not work.

One might think of compensating the removal of the Hilbert-space labels by the introduction of some other procedure. Indeed, in (10), in contrast to (11), the differential operator (repeated application of the momentum operator) containing the intrinsic parameter $m_{01}$ operates only on the single-particle function which contains the same parameter $m_{01}$, and the same holds for operator and wave function with $m_{02}$. Thus, we may indeed remove the Hilbert-space labels if we instead impose the explicit prescription that the operators operate only on those wave functions which have the same intrinsic parameters.

Compared with putting Hilbert-space labels at the variables and straightforwardly applying the usual rules of calculus, such an extra prescription is a rather clumsy procedure. It gets even clumsier if the interaction term in Eq.~(7) is taken into account, and still more so if the multi-particle wave function is no longer of product form but is written as $\Psi(x^{(1)},x^{(2)})$ or just $\Psi(1,2)$. In this latter case there are no longer any single-particle functions to be distinguished by means of intrinsic parameters. Of course, any multi-particle function can be expanded in a series of products of single-particle functions, but working out the expansion coefficients may be a laborious task, and putting Hilbert-space labels at the variables resolves the problem is a much more convenient way.

Thus, in systems of non-identical particles we might in principle do without Hilbert-space labels. These labels are only a convenient way of associating the single-particle operators with their respective single-particle wave functions. Operators and their associated wave functions must lie in the same Hilbert space. Note that we are here not concerned with endowing the physical quantum objects with additional properties, but only with constructing an appropriate formalism, a convenient book-keeping device.

Though the Hilbert-space labels have nothing to do with the mathematical form of the wavepackets it is
\begin{equation}
\psi(m_{01},u_1,x^{(1)},t)\;\psi(m_{02},u_2,x^{(2)},t)\;\not\equiv \psi(m_{01},u_1,x^{(2)},t)\;\psi(m_{02},u_2,x^{(1)},t)
\end{equation}
because the Hilbert spaces of wavepacket 1 and wavepacket 2, though isomorphic, are different [7, p.~1378, formula (B-3)]. Indeed, going from the left-hand side of formula (12) to the right-hand side is not innocuous because applying the Hamilton operator [in square brackets] of Eq.~(9) would evidently lead to an expression that is different from (10).

\bigskip
\noindent
{\textbf{5~~Self interactions}
\nopagebreak[4]
\smallskip

\noindent
Another reason for introducing Hilbert-space labels is to exclude self interactions. In the realist interpretation [1], as well as in de Broglie's conception of the wave function [12, p.~115, \S 50] the electron, say, is basically  an extended cloud of electric charge of density $e|\psi(\bfitr,t)|^2$. In this conception it cannot a priori be excluded that different regions of the cloud interact with each other via Coulomb forces. That is, the usual Schr\"odinger equation for the hydrogen atom, for example, should be complemented by an additional potential term of the form
\begin{equation}
\psi(\bfitr,t)\;e^2\,\int\frac{\psi^*\!(\bfitr',t)\psi(\bfitr',t)}{|\bfitr-\bfitr'|}\ud^3r',
\end{equation}
where $\bfitr$ and $\bfitr '$ denote different places within one and the same electron cloud. The Schr\"odinger equation with the additional term (13) is called the de Broglie equation by Tomonaga [12, p.~115, 315]. The term (13) has, however, to be omitted if the correct atomic energy levels are to be obtained. This was already noticed by Schr\"odinger in 1927 [13].

In quantum field theory the term (13) may be kept but has no effect because there the fields (wave functions) are turned into operators, the operators are expressed by means of one-particle creation and annihiliation operators, and due to ``normal ordering'' all annihilation operators operate first. In the term (13) two annihilation operators operate on one and the same one-particle state. That means that the second annihilation operator operates on the vacuum state, and this annihilates the term [12, p.~336 - 338], [14].

In the realist interpretation [1], which does not use concepts of relativistic quantum field theory, the exclusion of the term (13) is regarded as internal structurelessness of the wavepacket, mentioned in Section 2.

The Hilbert-space labels allow us to distinguish terms which describe the interaction between clouds representing different particles (cf. [7, p. 1426])
\begin{equation}
\int\frac{\varphi^*(\bfitr^{(1)},t)\:\eta(\bfitr^{(2)},t)}{|\bfitr^{(1)}-\bfitr^{(2)}|}\ud^3r^{(1)}\ud^3r^{(2)}
\end{equation}
from terms like (13), which describe the interaction between parts of one and the same cloud, and to exclude these latter terms.

\bigskip
\noindent
{\textbf{6~~Relabelling and renaming}}
\smallskip

\noindent
Relabelling, that is, permuting (exchanging, in the case of just two particles) the Hilbert-space labels leads to a different physical situation, as we have pointed out with formula (12).
On the other hand, relabelling means giving the wave functions other names, and physical description should be independent of the names chosen to denote the objects. In other words, physical description must be independent of renaming. What is the difference between relabelling and renaming?

In mathematical terms, leaving the physical description unchanged under relabelling means that the Schr\"odinger equation and all expressions of physical significance must be permutation invariant.
In quantum mechanics all expressions of physical significance can be expressed in terms of scalar products
\begin{equation}
(\Psi, \Phi).
\end{equation}
In these products $\Phi$ may also be written as $U\Phi'$ with $U=$ unitary operator of temporal evolution (=$\exp(-\rmi H_{\textrm{\footnotesize{int}}} t/\hbar)$, e.g.). $\Psi$ is taken at $t=0$, but $\Phi'$ at $t<0$ and by $U$ is brought to $t=0$.

The expectation value $\langle A\rangle=(\Psi,A\Psi)$ can be expressed by scalar products in the form $\langle A\rangle=\sum_i\lambda_i|(u_i,\Psi)|^2$, where $\lambda_i$ and $u_i$ are the eigenvalues and eigenfunctions, respectively, of the observable $A$. $(u_i,\Psi)$ can be taken as the amplitude of a transition from $\Psi$ to $u_i$ in a reduction (``measurement''). The wave function $\psi$ in the localization (action) probability $|\psi|^2\ud^3r$, in particular, is taken as the amplitude of a transition from $\Psi$ into a normalizable superposition of position eigenfunctions belonging to continuous eigenvalues. We may summarize these cases by considering the scalar product
\begin{equation}
(\Psi, T\Phi).
\end{equation}
Then, one requirement is that the product (16) must be permutation invariant under renaming. As the permutation operator $\Pal$ is unitary $(\Pal^{\dag}=\Pal^{-1})$ the $\Pal$ invariance of (16) is guaranteed if in addition to the wave function, the operator, too, is subject to a transformation by means of $\Pal$:
\begin{displaymath}
T \rightarrow \Pal T \Pal^{\dag} = \Pal T \Pal^{-1} \textrm{\  and \ } \Psi \rightarrow \Pal \Psi,\:\:   \Phi \rightarrow \Pal \Phi, \textrm{\quad so that\quad }
\end{displaymath}
\begin{displaymath}
(\Psi,T \Phi) \rightarrow (\Pal \Psi, \Pal T \Pal^{\dag}\Pal\Phi) =(\Pal^{\dag}\Pal\Psi,T \Pal^{\dag}\Pal\Phi) = (\Psi,T\Phi).
\end{displaymath}
This is just like the unitary transformations effecting a change of the basis in Hilbert space.

The transformation of both the operators and the wave functions also leaves the Schr\"odinger equation invariant:
\begin{displaymath}
\rmi \hbar\frac{\partial}{\partial t}\Psi=H\Psi\quad\rightarrow\quad\rmi\hbar\frac{\partial}{\partial t}\Pal\Psi=\Pal H\Pal^{\dag}\Pal\Psi=\Pal H\Psi,
\end{displaymath}
and multiplication from the left with $\Pal^{-1}$ yields
\begin{displaymath}
\rmi\hbar\Pal^{-1}\Pal\frac{\partial}{\partial t}\Psi=\Pal^{-1}\Pal H\Psi \textrm{\quad so that we return to \quad } \rmi\hbar\frac{\partial}{\partial t}\Psi= H \Psi.
\end{displaymath}
Thus, renaming means relabelling both the wave functions and the operators. This is not really surprising. Compare it with a police station which has a list containing the names Ike and Mike. Ike is to be imprisoned, and Mike is to be given a reward. If Ike and Mike exchange their names each of them will receive quite a different treatment, but if their names are exchanged in the police list, too, the situation will be as before.

\bigskip
\noindent
{\textbf{7~~Identical particles. Basic features}}
\smallskip

\noindent
Identical twins are two persons, but when we say that John William Strutt and Lord Rayleigh are identical we mean that they are one and the same person. So, the first lesson is that the word ``identical'' has several distinct meanings. These may be exemplified by pebbles, pingpong balls, and equal portions of water:

Pebbles can always be identified by their intrinsic properties such as size, mass, colour, shape and innumerable other properties revealed by more sophisticated methods. We might not be interested in distinguishing them, or we might, in certain situations, not be able to use more sophisticated methods (we may not be allowed to come sufficiently close) so that we cannot safely distinguish the pebbles, but in principle we always can.

Pingpong balls are equal in the sense that they all have the same intrinsic properties. Imagine that you are shown two pingpong balls lying on the table, one on the right-hand side and the other on the left-hand side. Then leave the room while the balls are moved by some other person. When you re-enter the room and see two balls on the table at the previous positions you cannot say whether the balls have been exchanged or not. If you had stayed in the room you could have followed their respective trajectories and you would indeed know whether they are exchanged or not.
Of course, if you are allowed to use every method of testing the balls you will always be able to discover intrinsic properties that are different in the two balls and can be used to identify them. This is not meant here. The pingpong balls conceived here have the same intrinsic properties under every test.

Equal water portions cannot even be identified by means of their trajectories. Imagine a black and a white espresso cup, each filled with 40~cm$^3$ of water. Pour the water of both cups into a jar, and then fill a red and a green cup with the water of the jar, giving each cup again 40~cm$^3$. In this case the question whether the water portion which now is in the red cup has come from the black or from the white cup does not make sense. Pebbles and pingpong balls are impenetrable, water portions are not.

Thus, pebbles can be identified by their intrinsic properties, pingpong balls by their trajectories, but equal water portions can only be identifided as long as they are kept apart from each other: these are different kinds of indistinguishability.

Where are these objects met in the description of nature? Pebbles are certainly met in classical (pre-quantum) mechanics of macroscopic objects. Whether our pingpong balls are met in classical mechanics is a debated question (Section 12). The atoms of early kinetic theory are certainly serious candidates. The atoms of quantum mechanics in their ground states may also be objects with the properties of our pingpong balls as long as they are localized at separate places as, for example, at the lattice sites of a crystal.
Objects with the properties of our water portions are met in quantum mechanics, where the objects are represented by wavepackets, which can overlap and condense (Sections 2, 13, 14).

Let us now turn to the description of identical particles in quantum mechanics. Here the characteristic feature of treating systems of identical particles is that it is based to a large extent on the treatment of non-identical particles: each wavepacket gets a Hilbert-space label.

In systems of non-identical particles the labels were in principle dispensable because their task could already be accomplished by the intrinsic parameters (mass $m_0$, charge $e$ etc.).
In systems of identical particles all intrinsic parameters are by definition equal, and now it is only the Hilbert-space labels at the variables $x, y, z$ that distinguish the particles in the formalism. Here the labels are not merely a convenient but an indispensible book-keeping device allowing to associate the one-particle operators with their respective one-particle wave functions.

Thus, the Schr\"odinger equation for the two electrons in the helium atom reads [7, p. 1419], [9, p. 257], [10, p. 690, 774]

\newcommand{\PP}{\emph{\boldmath $P$}}
\newcommand{\RR}{\emph{\boldmath $R$}}

\begin{equation}
\rmi\hbar\frac{\partial}{\partial t}\Psi=\left[\sum_{i=1}^2\frac{\PP _i^2}{2m_e}-\sum_{i=1}^2\frac{e^2}{|\RR _i|}+\sum_{i<j}^2\frac{e^2}{|\RR _i-\RR _j|}\right]\Psi
\end{equation}
with
\begin{equation}
\PP_i^2=-\hbar^2\left(\frac{\partial^2}{\partial (x^{(i)})^2}+\frac{\partial^2}{\partial (y^{(i)})^2}+\frac{\partial^2}{\partial (z^{(i)})^2}\right).
\end{equation}
$\RR_i$ is the vector pointing from the nucleus to a place in the cloud of electron $i$, and $m_e$ is the electron mass. The wave function $\Psi$ is antisymmetric 
\begin{equation}
\Psi(x^{(1)},x^{(2)})=- \Psi(x^{(2)},x^{(1)}).
\end{equation}
This property will be discussed in Section 11 below. For the present consideration the interesting feature is that the indices $i$ in Eq.~(18) are Hilbert-space labels, and that removing these labels from Eqs.~(18) and (19) would lead to wrong physical results, much as in the case of the hydrogen atom considered in Section 4, Eqs.~(9) and (11).

On the other hand, the labelling here introduces an ordering of the particles that has no support in any intrinsic physical properties of the particles. One might wish that in the description of systems of identical particles no ordinal numbers (number of your bank account) should appear, but only cardinal numbers (balance of your bank account), which specify how many particles belong to the system.

Indeed, a description of systems of identical particles, where only cardinal numbers are involved is the occupation-number or Fock representation. Fermions and bosons in this representation are distinguished by the possible values of the occupation numbers: 0, 1 for fermions, and 0,...,$\infty$ for bosons (elementary and $n$-fold condensed wavepackets, in the language of [1]). But this remains an ad hoc postulate unless it is deduced from the more general assumptions of wave mechanics, which lead to the Pauli exclusion principle and thus explain the occupation numbers.

Other approaches at getting along without Hilbert-space labels consist in working from the outset with a configuration space where coordinates which result from permutations do not define a new point [16], [15]. We do not discuss these approaches here.
The concern of the present note is only the usual text-book treatment of identical particles. In this treatment ordinal numbers, namely (Hilbert-space) labels, are introduced. But then, in order to remedy this, the ordering is neutralized by an additional procedure: permuting the labels in every wave function, and adding all functions with the permuted labels (the ``permutation'' functions, for short) to the original function. In this way we have labels and yet no ordering.

\bigskip
\noindent
{\textbf{8~~What is permuted?}}
\smallskip

\noindent
So far we have considered the exchange of the Hilbert-space labels. When the total wave function is given as a product
\begin{displaymath}
\Psi(1,2)=\psi(a_1,x_{01},t_{01},k_{01},x^{(1)})\;\psi(a_2,x_{02},t_{02},k_{02},x^{(2)})
\end{displaymath}
or as a superposition of products (which is possible for any wave function). The function parameter $\sigma$ is here replaced by (the set) $a$ because more general types of functions than just Gaussians are considered. One may then exchange the function parameters:
\begin{displaymath}
\Psi(2,1)=\psi(a_2,x_{02},t_{02},k_{02},x^{(1)})\;\psi(a_1,x_{01},t_{01},k_{01},x^{(2)})
\end{displaymath}
or exchange the Hilbert-space (particle) labels:
\begin{displaymath}
\Psi(2,1)=\psi(a_1,x_{01},t_{01},k_{01},x^{(2)})\;\psi(a_2,x_{02},t_{02},k_{02},x^{(1)}),
\end{displaymath}
and these two procedures are equivalent; they likewise remain so when the total permutation sum in case of more than two particles is considered. Both kinds of exchanges are met in the textbooks. Hilbert-space labels are permuted, for example, in [7, p.~1382] and [10, p.~587]. Function parameters (or, equivalently, function labels or function symbols) are permuted, for example, in [7, p.~208], [9, p.~364], [16, p.~333], [17, p.~100]. Dirac [6] discusses both procedures. This is essentially a question of notation, but it is worthwhile to pay it some attention.

There are many ways of writing a total wave function of product form:
\begin{equation}
\Psi(1,2)=\psi(a_1,x_{01},t_{01},k_{01},x^{(1)})\;\psi(a_2,x_{02},t_{02},k_{02},x^{(2)})
\end{equation}
\begin{equation}=\psi(u_1,x^{(1)}) \psi(u_2,x^{(2)})\textrm{\hspace{2.8cm}}
\end{equation}
\begin{equation}
=\psi^{(1)}(u_1)\psi^{(2)}(u_2)
\textrm{\hspace{3.6cm}}
\end{equation} 
\begin{equation}
= \psi_{u_1}(x^{(1)}) \psi_{u_2}(x^{(2)}) 
\textrm{\hspace{3.2cm}}
\end{equation}
\begin{equation}
=\varphi(x^{(1)}) \eta(x^{(2)})  
\textrm{\hspace{3.9cm}}
\end{equation}
\begin{equation}
=\varphi(1) \eta(2).
\textrm{\hspace{4.6cm}}
\end{equation}
The notation in formula (20) is like that in formula (1) but without the parameters $m_0$ and $t$ because they are the same for all particles. In formula (21) $u_1$ stands for the set $a_1, x_{01}, t_{01}, k_{01}$ of external parameters, which determine the shape and  position of the wavepacket, and the same for $u_2$ (cf. formula (3) in Section 3). In formula (22) the variables $x$ are suppressed and the Hilbert-space labels are put directly at the function symbols. Formula (23) is like formula (4), and formula (24) like formula (5). Note that we may only exchange either the Hilbert-space labels or the function parameters (labels, symbols). Exchanging both is the same as exchanging nothing.

In the general case the multi-particle wave function is no longer a single product of one-particle functions but a sum of such products. For two particles this reads
\begin{equation}
\Psi(1,2)=\sum_{r_1,r_2=1}^{\infty}c_{r_1r_2}\; \psi(u_{r_1},x^{(1)})\;\psi(u_{r_2},x^{(2)})=\sum_{r_1,r_2=1}^{\infty}c_{r_1r_2}\;\psi_{r_1}\!(x^{(1)})\;\psi_{r_2}\!(x^{(2)}) \textrm{\quad etc.}
\end{equation}
The exchange of Hilbert-space labels yields
\begin{displaymath}
\Psi(2,1)=\sum_{r_1,r_2=1}^{\infty}c_{r_1r_2}\;\psi(u_{r_1},x^{(2)})\;\psi(u_{r_2},x^{(1)}) \textrm{\quad etc.}
\end{displaymath}
and the exchange of the function parameters $u_r$ of the single-particle wave functions yields
\begin{displaymath}
\Psi(2,1)=\sum_{r_1,r_2=1}^{\infty}\;c_{r_1r_2}\;\psi(u_{r_2},x^{(1)})\;\psi(u_{r_1},x^{(2)}) \textrm{\quad etc.},
\end{displaymath}
and the result is the same.

Should the indices $r_i$ be exchanged both at the function parameters and at the multi-particle coefficients $c$
\begin{displaymath}
\Psi(2,1)=\sum_{r_2,r_1=1}^{\infty}c_{r_2r_1} \;\psi(u_{r_2},x^{(1)})\;\psi(u_{r_1},x^{(2)})
\end{displaymath}
the coefficients must be symmetric, $c_{r_1r_2}=c_{r_2r_1}$, if the exchange of the function parameters is to be equivalent with the exchange of the Hilbert-space labels.

However, while in standard quantum mechanics it is irrelevant whether we exchange (or permute) the Hilbert-space labels or the function parameters, our explanation of the spin-statistics connection in Section~11 shows that we have to make a choice, and that it is the exchange of the function parameters that we have to choose because among the parameters there is one whose exchange will give us the desired connection.

In no case does the exchange refer to an exchange of real physical particles. The exchange of the Hilbert-space labels attached to the position variable $x^{(1)}$ may have led to the misconception that the particles themselves had been exchanged, that is, that they had been transported from one place or quantum state to the other [18]. One easily gets trapped in this misconception if one pictures the quantum objects as pointlike: changing the position variable = changing the position of the object. Conceiving the quantum object to be an extended field structure described by a function of $\bfitr$, as in the realist interpretation, avoids that trap. In fact, even in the Copenhagen interpretation a particle shows up with a definite position only when the  position is subject to measurement. Thus, though nowadays physical particle transportation is no longer taken seriously, the language used in even excellent textbooks is still rather misleading [6], [7, p.~1377], [9, p.~364], [10, p.~585 - 588], [16, p.~335]. On the other hand, there are indeed books which conciously avoid speaking of particle exchange [19, p.~97], [20], [21].

The differing conceptions on the nature of the quantum object also have some bearing on the question of whether the spin component $m$ of the quantum object is to be counted among the parameters or the variables of the wave function. In Pauli's article [22] the spin component is counted among the variables. Pauli speaks of exchange of spin co-ordinates, position co-ordinates, particles, and labels as equivalent procedures, as is characteristic of the conception of point particles. The spin part of the wave function may be written as $\sigma_{\mu}(s_z)=\delta_{\mu s_z}$, where $\mu$ is a parameter of the wave function, namely an eigenvalue of the spin-component operator ($\equiv$ our $m$), and $s_z$ is the spin coordinate (cf. [17 p.~49, 50]).  But the Kronecker delta means that the two have always the same value. Actually, what really counts is that the scalar product satisfies $(\sigma_{\mu}(s_z),\sigma_{\mu'}(s_z))=\delta_{\mu\mu'}$. The spin coordinate $s_z$ serves only as a summation index when forming the scalar product in a special way.

In our conception the spin component $m$ is to be counted among the external parameters of the wave function.

\bigskip
\noindent
{\textbf{9~~Symmetric operators}}
\smallskip

\noindent
The requirement that every expression of physical significance be permutation invariant ($\Pal$ invariant, $\Pal$ symmetric) with respect to Hilbert-space or to function parameters can already be satisfied if only the operators are $\Pal$ invariant, the wave functions being $\Pal$ invariant or not, as already pointed out by Dieks [11]:

The $\Pal$ invariance of the operators $T$ means
\begin{displaymath}
\Pal T \Pal^{\dag}= T  \textrm{\quad or \quad}  [\Pal,T]=[T,\Pal]=0,
\end{displaymath}
and with this it follows that
\begin{displaymath}
(\Pal \Psi,T \Pal \Phi)=(\Psi,\Pal^{\dag}T\Pal\Phi)=(\Psi,T\Phi).
\end{displaymath}
The Schr\"odinger equation, too, is the same whether written with     $\Psi$  or with $\Pal \Psi$ [11]:
\begin{displaymath}
\rmi \hbar\frac{\partial}{\partial t}\Pal\Psi=H \Pal \Psi.
\end{displaymath}
Multiply from the left with $\Pal^{-1}$ and apply the preceding formulas with $T$ replaced by $H$
\begin{displaymath}
\Pal^{-1}\rmi \hbar\frac{\partial}{\partial t}\Pal \Psi= \Pal^{-1} H \Pal \Psi\;\rightarrow \;\rmi\hbar\frac{\partial}{\partial t}\Psi=H\Psi \textrm{\qquad q.e.d.}
\end{displaymath}
Postulating that in treating systems of identical particles the operators (observables), and in particular the Hamilton operator, should be permutation invariant is reasonable, as it is essentially the permutation invariance of the Hamilton function of classical meachanics.

Nevertheless, in spite of the fact that with $\Pal$ symmetric operators the  Schr\"odinger equation and every expression of physical significance is already $\Pal$ symmetric, current quantum mechanics postulates that the wave functions, too, be $\Pal$ symmetric. Why?

\bigskip
\noindent
{\textbf{10~~Symmetric wave functions}}
\nopagebreak
\smallskip

\noindent
Symmetric wave functions, to anticipate the answer to the question at the end of the preceding section, are postulated in order to remove the exchange degeneracy [7, Section XIV.C], [10, p. 585].

In quantum mechanics the general multi-particle wave function is not a product of well separated single-particle wave functions but a superposition of products of single-particle wave functions. That is, the single-particle wave functions (particles) may be entangled. This in turn means that the wave functions (particles) are (Bell) correlated, which does not occur with the particles in the classical treatment [23].

Consider two identical particles which are spatially well separated from each other and can therefore be represented by two non-overlapping (which includes: orthogonal) wave functions. The closest analogue to a classical description of these two particles is a wave function that is a product of the two single-particle wave functions. The two wave functions then may come together, overlap to some degree (with or without dynamical interaction), and finally two well separated wave functions may emerge again. Now, the rules of the game are such that each single-particle wave function must be given a label. Before the overlap let that particle which comes from the left have the label (1) and that from the right the label (2). The total wave function is $\Phi(x^{(1)},x^{(2)})=\xi(x^{(1)})\zeta(x^{(2)})$. This initial labelling is of course arbitrary. In the final state the single-particle functions must also be labelled, but in contrast to classical mechanics, in quantum mechanics the particles lose their individuality during the overlap,
 and in the final state, when the two wave functions are again well separated, nothing can tell us whether the labelling now has to be $\varphi(x^{(1)})\eta(x^{(2)})$ or $\varphi(x^{(2)})\eta(x^{(1)})$. The quantum objects are like the water portions conceived in Section 7.

As we presuppose that all observables commute with the permutation operator $\Pal$, all expressions of physical significance are $\Pal$ invariant, the wave functions being $\Pal$ symmetric or not (Section 9). In other words, we can describe all properties of the physical situation (physical state) just as well by means of $\varphi\!(x^{(1)})\eta\!(x^{(2)})$ as by means of $\varphi\!(x^{(2)})\eta\!(x^{(1)})$: the scalar product
\begin{equation}
S=\Big(\varphi\!(x^{(1)})\eta\!(x^{(2)}), O\!(x^{(1)},x^{(2)})\,\varphi\!(x^{(1)})\eta\!(x^{(2)})\Big)
\end{equation}
by (always allowed) renaming $x^{(1)}\leftrightarrow x^{(2)}$ becomes
\begin{displaymath}
S=\Big(\varphi\!(x^{(2)})\eta\!(x^{(1)}), O\!(x^{(2)},x^{(1)})\,\varphi\!(x^{(2)})\eta\!(x^{(1)})\Big)
\end{displaymath}
and by the symmetry $O\!(x^{(2)},x^{(1)})=O\!(x^{(1)},x^{(2)})$ this becomes
\begin{displaymath}
S=\Big(\varphi\!(x^{(2)})\eta\!(x^{(1)}), O\!(x^{(1)},x^{(2)})\,\varphi\!(x^{(2)})\eta\!(x^{(1)})\Big),
\end{displaymath}
and this is just the scalar product (27) with the exchange function $\varphi\!(x^{(2)})\eta\!(x^{(1)})$ instead of the original function $\varphi(x^{(1)})\eta(x^{(2)})$.  

The general superposition principle then demands that the physical situation is described just as well by the superposition
\begin{equation}
\Psi=\alpha\;\varphi(x^{(1)})\,\eta(x^{(2)})+\beta\;\varphi(x^{(2)})\,\eta(x^{(1)})
\end{equation}
with
\begin{equation}
|\alpha|^2 + |\beta|^2 = 1.
\end{equation}

Although the Hilbert-space labels have nothing to do with the mathematical form of the wavepackets, Eq.~(28) is not merely the superposition of two equal functions. The reason is that only the Hilbert-space labels are exchanged but not the function parameters (cf. (23), (24)). The operator operating on the variables $x^{(1)}$ meets a different function in the first than in the second term of the sum.

Eq.~(28) represents the well known exchange degeneracy [7, p.~1375 - 1377], [9, p.~365], [10, p.~583]. Physical predictions may now depend on the values of $\alpha$ and $\beta$. This is due to interference between the $\alpha$ and $\beta$ terms when expressions of physical significance are constructed which involve the superposition (28). To see this consider the scalar product
\begin{equation}
\left([\alpha\,\varphi\!(x^{(1)})\eta\!(x^{(2)})+\beta\,\varphi\!(x^{(2)})\eta\!(x^{(1)})],O\!(x^{(1)},x^{(2)})[\alpha\,\varphi\!(x^{(1)})\eta\!(x^{(2)})+\beta\,\varphi\!(x^{(2)})\eta\!(x^{(1)})]\right)
\end{equation}
which by renaming and using the symmetry of $O$ becomes
\begin{displaymath}
=\Big(\alpha\,\varphi\!(x^{(1)})\eta\!(x^{(2)}),O\alpha\,\varphi\!(x^{(1)})\eta\!(x^{(2)})\Big)+\Big(\alpha\,\varphi\!(x^{(1)})\eta\!(x^{(2)}),O\beta\,\varphi\!(x^{(2)})\eta\!(x^{(1)})\Big)
\end{displaymath}
\begin{displaymath}
+\Big(\beta\,\varphi\!(x^{(2)})\eta\!(x^{(1)}),O\alpha\,\varphi\!(x^{(1)})\eta\!(x^{(2)})\Big)+\Big(\beta\,\varphi\!(x^{(2)})\eta\!(x^{(1)}),O\beta\,\varphi\!(x^{(2)})\eta\!(x^{(1)})\Big)
\end{displaymath}
\begin{displaymath}
=(|\alpha|^2+|\beta|^2)\Big(\varphi\!(x^{(1)})\eta\!(x^{(2)}),O\varphi\!(x^{(1)})\eta\!(x^{(2)})\Big)
\end{displaymath}
\begin{equation}
+(\alpha^*\beta+\beta^*\alpha)\Big(\varphi\!(x^{(1)})\eta\!(x^{(2)}),O\varphi\!(x^{(2)})\eta\!(x^{(1)})\Big),
\end{equation}
where $(|\alpha|^2+|\beta|^2)=1$, but $(\alpha^*\beta+\beta^*\alpha)=2Re\{\alpha^*\beta\}$.

In order to get definite predictions (scalar products) the values of $\alpha$ and $\beta$ have to be fixed. As there are no a priori physical arguments in sight, this is done by way of a postulate. Of course, a basic requirement for the postulate is that it must lead to predictions that are confirmed by experiment. A postulate which succeeds in this and which fits well into the general symmetry requirements in systems of identical particles is
\begin{equation}
\alpha=\beta=\frac{1}{\sqrt{2}}
\end{equation}
resulting in a symmetric wave function (28), that is, a function that is invariant under the exchange of the Hilbert-space labels (1), (2) or the function parameters or the function symbols $\varphi,\eta$.

Actually in traditional quantum mechanics the postulate is not (32) but rather
\begin{equation}
\pm \alpha=\beta=\frac{1}{\sqrt{2}}.
\end{equation}
This will be discussed in the next section.

In the current formalism the total wave function of a system of identical particles is always assumed to be a superposition like (28) with (32) or (33). This amounts to assuming that the considered wave functions at some time in the past have overlapped and the entanglement thereby established has persisted. This has not led to any observed contradictions. Even if a superposition is formed both before and after the overlap this makes no difference to forming the superposition only before (or only after) the overlap because the symmetrizer $S=\sum_{\alpha}\Pal$ (and also the antisymmetrizer $\sum_{\alpha}\varepsilon_{\alpha}\Pal$) is a Hermitian projection operator (no longer unitary), so that (cf. [7, p. 1380])
\begin{displaymath}
(S\Psi,S\Phi)=(S^{\dag}S\Psi,\Phi)=(S^2\Psi,\Phi)=(S\Psi,\Phi).
\end{displaymath}
Only when there are explicit physical reasons to assume that the considered particles are independent (one particle being produced in the Andromeda galaxy and the other in our Milky-Way galaxy, say) the total wave function is taken as a simple product of one-particle wave functions.

Sometimes one finds the following ``proof'' of relation (33): the identity of the particles implies that the original and the exchange wave function must be essentially the same, that is, they can at most differ by a constant factor $\lambda$:
$\Psi(2,1)=\lambda\Psi(1,2)$.
Combining this with the defining equation of the permutation operator, $\Psi(2,1)=\Pal \Psi(1,2)$, we get $\Pal\Psi(1,2)= \lambda\Psi(1,2)$ so that $\Psi(1,2)$ must be an eigenfunction of $\Pal$. As (for any two particles) $\Pal^2=1$  the eigenvalues of $\Pal$ are $\pm 1$, so $\lambda=\pm 1$.
This is no proof but merely a different postulate because it is based on a requirement imposed directly on the $\Psi$ function, which in itself is no expression of direct physical significance.

\bigskip
\noindent
{\textbf{11~~Antisymmetric wave functions}}
\smallskip

\noindent
As described in Section~10, in conventional quantum mechanics the postulate for removing the exchange degeneracy is not Eq.~(32) but Eq.~(33), thus  admitting symmetric and antisymmetric wave functions. The justification for the two signs in Eq.~(33) in conventional quantum mechanics is based on the observation that if the wave function $\Psi$ is a solution of the Schr\"odinger equation with $\Pal$-symmetric Hamilton operator, then $\Pal\Psi$ is also a solution (cf. Section 9). $\Psi$ can then be written as a linear combination of functions grouped into systems that transform according to particular irreducible representations of the permutation group. The postulate (33) then is equivalent to the postulate that only the two representations of degree (dimension) 1 are relevant for the description of nature [10, p.~1117], [22]. The wave function (28) with the plus sign from Eq.~(33), meaning bosons, is empirically found to represent particles with integral spin, in units of $\hbar$, and the function with the minus sign, meaning fermions, is found to represent particles with half-integral (i.e. half-odd-integral) spin. This connection between spin and statistics was first proved by Fierz and by Pauli in the framework of relativistic quantum field theory [24]. In the last 20 years a rising number of papers with a peak statistical frequency in 2004 have appeared which tried to prove such a spin-statistics connection under various sets of conditions, in particular within the framework of quantum mechnics.

I will here describe my own proposal [25]. It follows the suggestion by Feynman [26], that the exchange degeneracy is fixed by Eq.~(32), that is, by the plus sign only. The minus sign arises in the construction of the exchange function from the original function. The single-particle wave functions are of course complemented by a spin part. The spin-component eigenfunctions depend on the quantum number $m$ and on the azimuthal angle $\chi$ of a rotation about the spin-quantization axis in the form $\exp(\rmi m\chi)$. This is standard quantum mechanics [7, p.~703, 984, 985]. In standard quantum mechanics the angle $\chi$ is however considered to have no physical significance, and is usually neglected. 
In my approach the external parameter $\chi$ is exchanged along with the other parameters, and thereby acquires physical significance because it becomes a relative phase angle. In the case of half-integral $m$ for each value of $\chi$ there are two eigenfunctions, which differ by their signs. This is the well known spinor ambiguity, and like the ambiguity represented by the exchange degeneracy, it has also to be fixed. This is done by effecting the exchange of
$\chi$ by way of rotation, and admitting only rotations in one sense, either clockwise or counterclockwise. If it is done so, the second term on the right-hand side of Eq.~(28) acquires the factor $(-1)^{2s}$, $\hbar s$ being the total spin of the particle. 

The essential step leading to this result can be outlined in a simple example: start from the special function \quad $\psi^{(1)}(u_a,\chi_a)\psi^{(2)}(u_b,\chi_b)$, \quad where both single-particle functions belong to the same $m$, and $\chi_b > \chi_a$. The exchange of the parameters $u_a$ and $u_b$ changes the function into \quad$\psi^{(1)}(u_b,\chi_a)\psi^{(2)}(u_a,\chi_b)$.\quad Then, when $\psi^{(1)}(u_b,\chi_a)$ is rotated counterclockwise from	$\chi_a$ to $\chi_b$ we have \quad$\psi^{(1)}(u_b,\chi_b)=\exp(\rmi m(\chi_b-\chi_a))\psi^{(1)}(u_b,\chi_a)$.\quad Likewise, when \quad$\psi^{(2)}(u_a,\chi_b)$ \quad is rotated counterclockwise from $\chi_b$ to $\chi_a$ we have \quad$\psi^{(2)}(u_a,\chi_a)=\exp(\rmi m(2\pi+\chi_a-\chi_b))\psi^{(2)}(u_a,\chi_b)$. \enskip In this way the product\enskip $\psi^{(1)}(u_b,\chi_a)\psi^{(2)}(u_a,\chi_b)$ becomes\enskip $F\times \psi^{(1)}(u_b,\chi_b)\psi^{(2)}(u_a,\chi_a)$, where 
\begin{equation}
F=\exp(-\rmi m(\chi_b-\chi_a))\exp(-\rmi m(2\pi+\chi_a-\chi_b))=(-1)^{2m}=(-1)^{2s}.
\end{equation}

It is seen that this result can only be obtained when the function parameters, not the Hilbert-space labels, are exchanged. Exchange of function parameters and exchange of Hilbert-space labels are here not equivalent. For the described procedure the multi-particle wave function $\Psi(1,2,\ldots,N)$ must be given in terms of one-particle functions, but this is in principle always possible (cf. Eq.~(26) in Section 8).

The result (37) also holds for functions of general form, for any number of particles, and in the relativistic domain. The connection between spin and statistics is thus changed from an axiom into a theorem, revealing a larger domain of phenomena which good old quantum mechanics can         incorporate into a coherent theoretical picture. The antisymmetric wave functions do not emerge from the permutation group with its two inequivalent one-dimensional representations, but from the rotation group with the ambiguity of its spinor representation. This entails that only Bose and Fermi statistics, but no parastatistics result.

The antisymmetric wave functions lead to the Pauli exclusion principle, which may be formulated by saying that ``no antisymmetric wavefunction describing an N-electron atom can be made up of one-electron functions, two of which have the same set of one-electron quantum numbers'' [17, p.~101]. Or: no antisymmetric wave function exists which could describe two fermions represented by two equal one-particle wave functions (cf. [7, p. 1389]).

\bigskip
\noindent
{\textbf{12~~The term $\emph{\boldmath{$-k\ln N!$}}$ in the entropy}}
\smallskip

\noindent
We will now consider some implications of the preceding considerations for statistical mechanics.
Either term on the right-hand side of formula (28) is a product of single-particle wave functions and is thus the closest quantum mechanical analogue to a classical situation: the first term would mean that particle 1 is in phase-space region $a$ and particle 2 in region $b$, whereas the second term would mean the reverse. In classical statistical mechanics these two states are considered as different. In other words, exchanging the particles leads to a new classical state. In quantum mechanics there is only one state, given by formula (28), and with the choice (32) or (33) of the coefficients $\alpha$ and $\beta$ the exchange of Hilbert-space labels does not lead to a new state. In the case of $N$ identical particles there are $N!$ possible classical states, according to the number of possible permutations of $N$ elements, but still only one quantum mechanical state. In other words, when going from classsical to quantum mechanics the number of states has to be divided by $N!$ This is a justification of the well-known subtraction of the term $k \ln N!$ in the entropy formula of classical statistical mechanics in order to resolve the Gibbs paradox concerning the entropy of mixing [19, p.~193], [27, p.~141], [28].	Notice that this holds independently of whether (32) or (33) is chosen, that is, independent of whether the quantum mechanical states are symmetric or antisymmetric, i.e. describe bosons or fermions.

Whether quantum mechanics is really necessary for the justification of the division by $N!$ is a debated question [19, p.~96], [29]. It depends on whether or not one counts the pingpong balls of Section 7 among the classical objects. If one attributes the properties of our pingpong balls to the classical particles, it is true that the various ways of filling or refilling the sites with particles can be distinguished, but when only the resulting arrangement is seen it is no longer possible to distinguish the various ways that have led to it. The entropy is a function of state, that is, it depends only on the resulting arrangement and not on its past history. Then division by $N!$ makes already sense in classical statistical mechanics.

\bigskip
\noindent
\textbf{13~~Bose, Fermi and Boltzmann probabilities}}
\nopagebreak
\smallskip

\noindent
The essential new feature of quantum as compared to classical particles is that there are situations where quantum particles cannot even be identified by their position. This happens when there is, at some time, some overlap of the wavepackets which represent the particles. In particular, in quantum mechanics it can happen that two or $n$ single-particle wave functions after the overlap are equal. This, in the realist language of [1], means $n$-fold condensed wavepackets, and in the language of statistical mechanics it means a multiple occupation of quantum phase-space cells of size $h^3$. The objects within such a quantum cell can no longer be distinguished by any means, and permutation of them does not even make sense. They are comparable to water portions in a jar.

In the Fermi case a total wave function with two or more single-particle wave functions being equal is excluded by the appearance of the minus sign in formula (33), leading to the Pauli exclusion principle. Accordingly, never more than two equal fermions can be in one quantum cell, and it might seem that there is never any overlap between fermion functions, so that fermions are like pebbles and not like water portions. However, this refers to the situation after the process of entanglement. Before this process had set in, there has been overlap because the general single-particle wave function is a superposition of several spin-component eigenfunctions. It is easy to convince oneself that the scalar product between such superpositions is nonzero (they are not orthogonal), and that the nonzero terms exhibit spatial overlap between one-particle functions which have the same spin component. Only if the total state initially is a single product of one-particle functions, and the spin components are all different, can the particles be considered as effectively distinguishable, i.e., as pebbles (cf. [25, Section~7]).

Classical statistical mechanics considers the division of phase space into distinguishable regions (energy regions, cells, groups, levels, shells etc. in the various textbooks) but does not consider the subdivision of the chosen regions into quantum cells of size $h^3$ (subcells, compartments, quantum states etc.). In quantum mechanics this subdivision is instrumental in determining the probability $w_i$ of having $n_i$ particles in a region $i$ comprising $g_i$ quantum cells and leads to the well-known formulas for bosons
\begin{equation}
w_i=\frac{(n_i+g_i-1)!}{n_i!(g_i-1)!}= \left(\!\!\begin{array}{c}n_i+g_i-1\\n_i\end{array}\!\!\right)=\left(\!\!\begin{array}{c}n_i+g_i-1\\g_i-1\end{array}\!\!\right)
\end{equation}
and for fermions
\begin{equation}
w_i=\frac{g_i!}{n_i!(g_i-n_i)!}=\left(\!\!\begin{array}{c}g_i\\n_i\end{array}\!\!\right).
\end{equation}
As stated above, permutation of the particles within one quantum cell does not make sense and so should not even be mentioned. Actually, the expression $n_i!$, which means permutation of all particles in region $i$, even those that might be in one and the same quantum cell, appears in both formulas. But this is only so in the mathematical formula and does not appear in the definition of the $w_i$: in the Fermi case $w_i$ is defined as the number of ways in which $n_i$ quantum cells can be chosen from $g_i$ distinguished quantum cells into which the $n_i$ particles can be put. In the Bose case $w_i$ is defined as the number of ways in which the number $n_i$ can be decomposed into the sum of $g_i$ nonnegative integers, namely the respective number of particles put in each of the quantum cells [19, p. 70, 100].

In the limit $n_i\ll g_i$ the Bose and the Fermi formulas tend to 
\begin{equation}
\frac{g_i^{n_i}}{n_i}\left(1\pm \frac{n_i(n_i-1)}{2g_i}\right),
\end{equation}
where the upper sign refers to bosons and the lower sign to fermions. In the Bose case formula (40) can be obtained by writing
\begin{displaymath}
\frac{(n_i+g_i-1)!}{n_i!(g_i-1)!}=\frac{g_i^{n_i}}{n_i!}\;(1+1/g_i)\,(1+2/g_i)\cdots (1+(n_i-1)/g_i)
\end{displaymath}
and neglecting all terms of second and higher order in $(1/g_i)$ when evaluating the product. The Fermi case obtains in an analogous way.

Thus in the limit $n_i/g_i\rightarrow 0$ both the Bose and the Fermi formula tend to
\begin{equation}
w_i=\frac{g_i^{n_i}}{n_i!},
\end{equation}
which may be called the correct Boltzmann formula [27, p.~182]. The limit $n_i/g_i\rightarrow 0$ means that there are many more quantum cells than particles in region $i$, so that the probability of two or more particles occupying the same quantum cell becomes zero. This is the classical limit because in this limit the size $h^3$ of a quantum cell goes to zero and the number $g_i$ of quantum calls in the region $i$ goes to infinity.

The approach from the quantum to the classical case can also be seen from the superposition (28) when it is generalized to comprise sums of products of $n$ one-particle wave functions
\begin{equation}
\Psi=\sum_{r_1,r_2,\cdots, r_n=1}^g c_{r_1r_2\cdots r_N}\;\psi_{r_1}(x^{(1)})\;\psi_{r_2}(x^{(2)})\cdots\psi_{r_n}(x^{(n)}),
\end{equation}
where the symmetry type of the coefficients may be left open. $n$ is the number of actually present wavepackets (particles), $g$ is the (maximum) number of formally possible elementary (one-quantum) wavepackets in the chosen energy interval.

The sum (42) has $g^n$ terms. The number of terms which have all indices different is
\begin{footnotesize}
$\left(\!\!\begin{array}{c}g\\n\end{array}\!\!\right)\!\!$
\end{footnotesize}
$n!$, namely the number 
\begin{footnotesize}
$\left(\!\!\begin{array}{c}g\\n\end{array}\!\!\right)$
\end{footnotesize}
of possibilities of choosing $n$ different values out of the $g$ $(\ge n)$ possible values of the indices times the number $n!$ of permutations of the $n$ chosen values among themselves. The fraction of terms with two or more equal indices is therefore
\begin{equation}
1-\frac{g!}{(g-n)!g^n}\rightarrow \frac{n(n-1)}{2g}\rightarrow 0
\end{equation}
when 	$g\rightarrow \infty$, $n$ fixed. This can be seen by a consideration like that leading to formula (40).

Thus, whatever the values of the coefficients $c_{r_1r_2\cdots r_n}$ in the terms with two or more equal indices, the contribution of these terms to the sum (42) sinks into insignificance compared with the overwhelming number of the other terms when $n/g\rightarrow 0$. This is independent of whether $\Psi$ in (42) is symmetrized or not because the described considerations would apply to each of the terms with permuted labels in the symmetrized $\Psi$.

\bigskip
\noindent
{\textbf{14~~The Bose and Fermi distributions}}
\nopagebreak
\smallskip

\noindent
By Bose and Fermi distributions I mean
\begin{equation}
N\ud p=\frac{g_p}{\exp[(\varepsilon-\mu)/kT]\pm 1},
\end{equation}
where $N\ud p$ means, in the usual interpretation, the time averaged number of particles in some volume $V$ whose absolute value of momentum lies in the interval $\ud p$ about $p$. $\mu$ is the chemical potential. $g_p=4\pi V p^2 \ud p/h^3$, and $\varepsilon=[p^2c^2+(mc^2)^2]^{1/2}$ is the total energy of a particle. The distributions (44) can be obtained from the formulas (38) and (39) through the well known procedure of maximizing the entropy $S=k \ln \prod_i w_i$.

I would like to mention here that the distributions can also be obtained by considering $n$-fold condensed wavepackets (wavepackets representing $n$ quanta) as the fundamental units of an ideal gas. $g_p$ is then the maximum possible number of wavepackets in $V$ covering the momentum interval $\ud p$. The packets can be treated as statistically independent objects, whereas the quanta, if we were to take these as entities of their own, are not independent. Actually, what in the usual interpretation of formula (44) is the number of particles, is in the realist interpretation the number of quanta. For when a wavepacket causes an effect, a response of a counting apparatus say, this is usually interpreted as the count of a particle. Thus the statistical interdependence of the quanta shows up as the ``mutual influence of the molecules which for the time being is of a quite mysterious nature'' [23].
The wavepackets can condense, decondense, and exchange energy among themselves. This leads to a simple and symmetric balance equation, following Einstein's method of emission and absorption of radiation by atoms [30]:
\newcommand{\rmf}{\mathrm{f}}
\begin{equation}
p(s,\varepsilon_1^{\rmi})p(r,\varepsilon_1^{\rmf})q(s',\varepsilon_2^{\rmi})q(r',\varepsilon_2^{\rmf})=p(s-n,\varepsilon_1^{\rmi})p(r+n,\varepsilon_1^{\rmf})q(s'-n',\varepsilon_2^{\rmi})q(r'+n',\varepsilon_2^{\rmf}).
\end{equation}
Two kinds of wavepackets are considered (Einstein: atoms and photons). $p(s,\varepsilon_1^{\rmi})\ud\varepsilon_1^{\rmi}$ is the mean number of $s$-fold condensed packets of kind 1 in volume $V$ that represent $s$ quanta in the energy interval $\ud\varepsilon [=(\partial\varepsilon / \partial p)\ud p]$, and analogously with the other terms. $q(\,)$ refers to packets of kind 2. The solution of Eq.~(45), when energy conservation in the form $n(\varepsilon_1^{\rmi}-\varepsilon_1^{\rmf})=n'(\varepsilon_2^{\rmf}-\varepsilon_2^{\rmi})$ is taken into account, is
\begin{equation}
p(s,\varepsilon)=a(\varepsilon) \exp[-(b\varepsilon-c)s]
\end{equation}
\begin{equation}
q(s,\varepsilon)=a'(\varepsilon)\exp[-(b\varepsilon-c')s],
\end{equation}
where $b=1/kT, c=\mu/kT, c'=\mu'/kT$ follow from thermodynamic relations involving the entropy
\begin{equation}
S=k\ln \prod_{\{d\varepsilon_i\}}\frac{g_p!}{[p(0,\varepsilon_i)d\varepsilon_i]!\;[p(1,\varepsilon_i)d\varepsilon_i]! \cdots}.
\end{equation}
Then, when observing that $\sum_{\{s\}}p(s,\varepsilon)\ud\varepsilon=g_p$ and that for Bose packets the numbers $s, r$ etc. in Eqs.~(45) to (47) are nonnetative integers, while for Fermions they can only be 0 or 1, the total number of quanta
\begin{displaymath}
N\ud p=\sum_{\{s\}}s p(s,\varepsilon)d\varepsilon
\end{displaymath}
is given by the formulas (44). And as I have said, these quanta, not the wavepackets, are identified with the particles in the usual interpretation of formula (44). The usual fluctuation formulas for quantum counts can be obtained on the same line.

\bigskip
\noindent
{\textbf{References}}
\begin{enumerate}
\renewcommand{\labelenumi}{[\arabic{enumi}]}
\begin{sloppypar}

\item Jabs, A.: Quantum mechanics in terms of realism, arXiv:quant-ph/9606017 

\item Jabs, A.: An Interpretation of the Formalism of Quantum Mechanics in Terms of Epistemological Realism, arXiv:1212.4687 (Brit. J. Philos Sci. \textbf{43}, 405 - 421 (1992))

\item Piaget, J.: \emph{The Construction of Reality in the Child} (Routledge \& Kegan Paul, London, 1954, translation from the French) p. 4

\item Glasersfeld, E. von: \emph{Radical Constructivism. A Way of Knowing and Learning} (The Falmer Press, London, 1995); Foerster, H. von: \emph{Observing Systems} (Intersystems Publications, Seaside CA, 1982)

\item Andersen, H.C.: The Emperor's New Clothes, http://www.andersen.sdu.dk/
\\vaerk/hersholt/TheEmperorsNewClothes\_e.html (J. Hersholt, \emph{The Complete Andersen} (The Limited Editions, New Club, New York, 1949, translation from the Danish)). -- And what did the emperor and his noblemen do after the whole town joined in the child's words? ``So, he walked more proudly than ever, as his noblemen held high the train that wasn't there at all'' 

\item Dirac, P.A.M.: \emph{The Principles of Quantum Mechanics} (Oxford University Press, London, 1967) p.~207, 217, 218

\item Cohen-Tannoudji, C., Diu, B., Lalo\"e, F.: \emph{Quantum Mechanics}, Vols. I, II (John Wiley \& Sons, New York, 1977, translation from the French)

\item Wohl, C.G., et al.: Review of Particle Properties, Rev. Mod Phys. \textbf{56} (2), Part II (1984)

\item Schiff, L.I.: \emph{Quantum Mechanics} (McGraw-Hill, New York, 1968)

\item Messiah, A.: \emph{Quantum Mechanics}, Vols.~1,~2 (North-Holland, Amsterdam, 1970, 1972, translation from the French)

\item Dieks, D.: Quantum Statistics, Identical Particles and Correlations, Synthese \textbf{82}, 127 - 155 (1990)

\item Tomonaga, S.-I.: \emph{Quantum Mechanics}, Vol. II (North-Holland Publishing Company, Amsterdam, 1966, translation from the Japanese)

\item Schr\"odinger, E.: Annalen der Physik (Leipzig) \textbf{82} (4), 265 - 273 (1927). English translation in: \emph{Collected Papers on Wave Mechanics} (Chelsea Publishing Company, New York, 1982) p. 136

\item D\"oring, W.: \emph{Atomphysik und Quantenmechanik}, Vol.~II  (de Gruyter, Berlin, 1976) p. 402

\item Laidlaw, M.G.C., DeWitt, C.M.: Feynman Functional integrals for Systems of Indistinguishable Particles, Phys. Rev. D \textbf{3} (6), 1375 - 1378 (1971); Nelson, E.: \emph{Quantum Fluctuations} (Princeton University Press, Princeton, 1985) p. 100 - 102; Leinaas, J.M., Myrheim, J.: On the Theory of Identical Particles, Nuovo Cimento \textbf{37B} (1), 1 - 23 (1977)

\item Gottfried, K.: \emph{Quantum Mechanics} (Benjamin, Reading MA, 1966)

\item Cowan, R.D.: \emph{The Theory of Atomic Structure and Spectra} (University of California Press, Berkeley, 1981)

\item Heisenberg, W.: Mehrk\"orperproblem und Resonanz in der Quantenmechanik, Z. Phys. \textbf{38}, 411 - 426 (1926), especially p. 421

\item Ter Haar, D.: \emph{Elements of Thermostatistics} (Holt, Rinehart and Winston, New York, 1966)

\item Peres, A.: \emph{Quantum Theory: Concepts and Methods} (Kluwer, Dordrecht, 1993) p. 126

\item Davydov, A.S.: \emph{Quantum Mechanics}, 2nd edition (Pergamon Press, Oxford, 1976, translation from the Russian) p. 306

\item Pauli, W.: Die allgemeinen Prinzipien der Wellenmechanik, in: \emph{Handbuch der Physik}, Vol. 24, Part I, edited by Geiger, H., Scheel, K. (Springer-Verlag, Berlin, 1933) Chapter 2, Number 14. English translation of a 1958 reprint: \emph{General Principles of Quantum Mechanics} (Springer-Verlag, Berlin, 1980) Chapter VII, Section 14(a)

\item Einstein, A.: Quantentheorie des idealen einatomigen Gases, Zweite Abhandlung, Sitzungsberichte der Preu\ss ischen Akademie der Wissenschaften, Physikalisch-Mathematische Klasse, Berlin, 1925, p. 3 - 14

\item Fierz, M.: \"Uber die relativistische Theorie kr\"aftefreier Teilchen mit beliebigem Spin, Helv. Phys. Act. \textbf{12}, 3 - 37 (1939); W. Pauli, The connection between spin and statistics, Phys. Rev. \textbf{58}, 716 - 722 (1940)

\item Jabs, A.: Connecting spin and statistics in quantum mechanics, arXiv:\linebreak0810.2399

\item Feynman, R.P.: The reason for antiparticles, in: \emph{Elementary Particles and the Laws of Physics}, edited by Feynman, R.P., Weinberg, S. (Cambridge University Press, New York, 1987) p. 1 - 59, especially p. 56 - 59

\item Huang, K.: \emph{Statistical Mechanics}, 2nd edition (John Wiley \& Sons, New York, 1987)

\item Becker, R.: \emph{Theorie der W\"arme} (Springer-Verlag, Berlin, 1961) p. 24, 25, 117, 118. English translation: \emph{Theory of Heat} (Springer-Verlag, Berlin, 1967) p.~26, 134, 135

\item Saunders, S.: On the Explanation for Quantum Statistics, arXiv:quant-ph/0511136 (Studies in the History and Philosophy of Modern Physics \textbf{37}, 192 - 211 (2006))

\item Einstein, A.: Zur Quantentheorie der Strahlung, Physik. Zeitschr. \textbf{18}, 121 - 128 (1917). English translation in: Ter Haar, D.: The old quantum theory (Pergamon Press, Oxford, 1967)

\end{sloppypar}
\end{enumerate}
\bigskip
\hspace{5cm}
------------------------------
\end{document}